\documentclass[11pt,a4paper]{article}
\pdfoutput=1
\usepackage{jcappub}
\usepackage{graphicx}
\usepackage{amsmath}   
\usepackage{amssymb}

\def\eq#1{{Eq.~(\ref{#1})}}

\title{Intergalactic Lyman-$\alpha$ haloes before reionization are detectable with JWST}

\author[a]{Hamsa Padmanabhan,}

\author[b]{Abraham Loeb}

\affiliation[a]{D\'epartement de Physique Th\'eorique, Universit\'e de Gen\`eve \\
24 quai Ernest-Ansermet, CH 1211 Gen\`eve 4, Switzerland\\}

\affiliation[b]{Astronomy department, Harvard University \\
60 Garden Street, Cambridge, MA 02138, USA}

\emailAdd{hamsa.padmanabhan@unige.ch}
\emailAdd{aloeb@cfa.harvard.edu}

\abstract{
The {\it James Webb Space Telescope (JWST)} recently reported a large population of UV luminous galaxies at high redshifts, $ z > 10$, as well as Lyman-$\alpha$ emitting (LAE) galaxies out to $z \sim 11$. We use the observed UV luminosities along with a data-driven approach at lower redshifts to place constraints on the observability of the intergalactic Lyman-$\alpha$ intensity, scattered in the form of Loeb-Rybicki haloes, during the pre-reionization and reionization epochs ($z \sim 9-16$). We forecast the sensitivity and resolution required to detect these intergalactic haloes, finding that individual haloes with LAE luminosities $> 10^{43}$ ergs/s are detectable at a few sigma level at $z \lesssim 9$, while stacking of  $\sim 10$ haloes is expected to result in detections out to $z \sim 16$. Finding these haloes is expected to shed light on the neutral intergalactic hydrogen during cosmic reionization.
}

\begin{document}
\maketitle
\flushbottom

\section{Introduction}

Cosmic reionization, the transition when the bulk of the intergalactic medium (IGM)  changed from being primarily neutral to largely ionized,  is a key milestone in the evolution of the Universe \citep[e.g.,][]{loeb2013}. Recent observations of the transmitted flux in the spectra of quasars have indicated reionization to be nearly complete by $z \sim 5.5-5.7$ (e.g., \citep{bosman2022}), while the Thomson scattering optical depth of the Cosmic Microwave Background (CMB) is consistent with the midpoint of reionization occurring around $z \sim 7.7$ \citep{planck2020}.

Current evidence indicates star-forming galaxies to be the major contributors to cosmic reionization \citep[e.g.,][]{bunker2004, Bouwens2015b, Finkelstein2019}. Of these,  the Lyman-Alpha Emitters (LAEs, \cite{hu1996, steidel1996}) are  important probes of the neutral hydrogen fraction in the IGM  \citep{Zheng-2017, Mason-2019, ouchi2020, Umeda-2023, Nakane-2023} in the later stages of reionization, $z \lesssim 7$.
The {\it James Webb Space Telescope (JWST)} NIRSpec and NIRCam instruments have  been successful in observing Lyman-$\alpha$ emitting galaxies out to $z \sim 11$ \citep{Bunker-2023, Tacchella-2023, Jung2023, Tang-2024, Saxena2023, whitler2024, Witstok-2024, witten2024}. Recently, a  large number of UV-luminous galaxies have also been detected by JWST \citep[e.g.,][]{donnan2023, harikane2023} leading to the development of UV luminosity functions over $z \sim 9-16$, and complementing those found at lower redshifts from the Hubble Space Telescope \citep[HST;][]{bouwens2015}.

At at $z > 6$, the Lyman-$\alpha$ line  becomes increasingly difficult to observe in the IGM, due to the Gunn-Peterson effect \citep{gunnpeterson} that suppresses its transmission at these early times (for a recent review, see, e.g. \citep{fan2023}).  Since the line is resonant, Lyman-$\alpha$ radiation is typically re-emitted (scattered) as soon as it is absorbed by a hydrogen atom, leading to the diffusion of Lyman-$\alpha$ photons through the medium until they are either destroyed by dust,  or redshift out of the resonance and escape due to the Hubble expansion of the medium \citep[for a review, see, e.g.,][]{dijkstra2014}. Around a Lyman-$\alpha$ emitting galaxy, this behaviour leads to the development of so-called Loeb-Rybicki Lyman-$\alpha$ haloes \citep{loeb1999}, which are several arcminute scale features observable in the neutral IGM expanding cosmologically around a steady point source. Such haloes may be observable  up to the pre-reionization epoch ($z \sim 10-20$), where detection of the { Gunn-Peterson halo is much more difficult since it is scattered from a much larger and dimmer region than the Lyman-$\alpha$ halo,  due to the longer distance travelled by the photons originating blueward of the Lyman-$\alpha$ resonance \citep{loeb1999}}.

In this paper, we use the Lyman-$\alpha$ emitter luminosity functions compiled at $z \sim 3-7$ from the Subaru telescope surveys to describe the evolution of Lyman-$\alpha$ luminosity with dark matter halo mass, which we then compare to their intrinsic Lyman-$\alpha$ luminosities inferred from the UV luminosity functions observed by the HST at the same epochs.  From these two datasets,  we derive the expected evolution of the escape fraction of LAEs, finding it to evolve consistently with recent as well as older observations. The data suggest a Lyman-$\alpha$ escape fraction of 0.1-0.2 at the higher redshift range, $z \sim 9-16$. We use this result, along with the UV luminosity functions compiled over the same redshift range by the JWST NIRCam surveys, to infer the expected Lyman-$\alpha$ luminosities of LAEs at $z \sim 9-16$ as a function of their dark matter halo masses (although these galaxies are not expected to be directly observable due to the increasing neutral fraction in the IGM). We predict the intensity of scattered radiation around these sources --  the Loeb-Rybicki haloes -- and forecast the required sensitivity of their detection with JWST. We find that at the highest luminosities, individual haloes should be detectable up to { $z \lesssim 9$}, and the stacking of $\sim$ 10 LAE  should lead to their detectability out to $z \sim 16$, with spectral resolution requirements of $R \sim 200-500$.

The paper is organized as follows. In Sec. \ref{sec:abmatch}, we describe the theoretical formalism relating the Lyman-$\alpha$ luminosities of LAEs over $z \sim 3-7$ to the masses of their host dark matter haloes. We validate the procedure by evaluating the clustering of Lyman-$\alpha$ haloes across redshifts, as compared to recent observations. We then, in Sec. \ref{sec:loebrybicki}, use the results so derived to forecast the observed intensity of Loeb-Rybicki haloes at $z > 10$, and formulate the resolution and sensitivity requirements for their detectability with JWST. We summarize our conclusions and discuss future prospects in Sec. \ref{sec:conclusions}. Throughout the paper, we use a $\Lambda$CDM cosmology with matter density parameter $\Omega_m = 0.27$, dark energy density parameter $\Omega_{\Lambda} = 0.73$, baryon density parameter $\Omega_b = 0.042$, and the present-day Hubble parameter $H_0 = 100 \ h \ $ km/s with $h = 0.71$.

\section{Abundance matching and comparison to low-redshift results}
\label{sec:abmatch}

\begin{figure}
    \centering
    \includegraphics[width = 0.45\columnwidth]{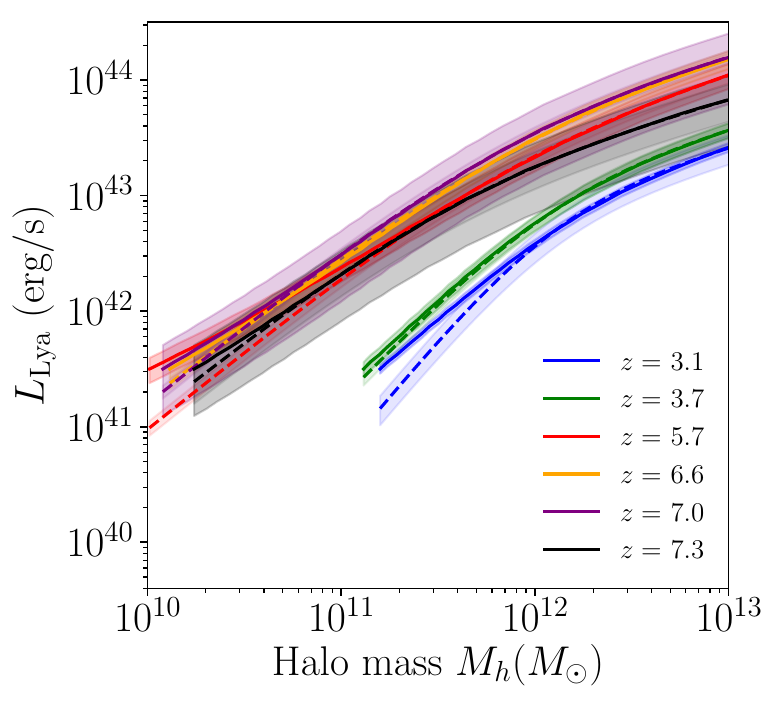} 
     \includegraphics[scale =0.45]{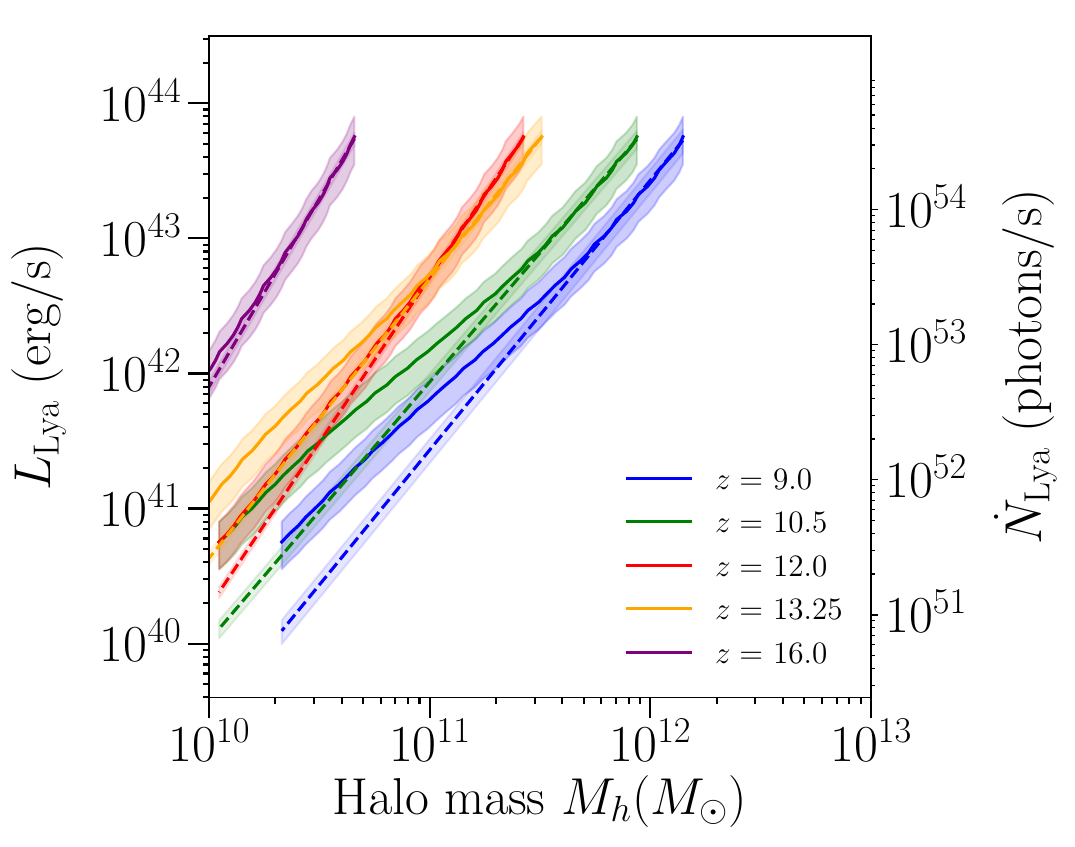} 
    \caption{{\it Left panel:} Results of abundance matching the observed Lyman-$\alpha$ luminosity functions over $z \sim 3-7$. The dashed lines show the abundance matching results, and the solid ones the fitting forms outlined in Table \ref{table:constraints}, with {  their enclosing shaded regions denoting the uncertainties inferred from the Lyman-$\alpha$ luminosity function and the fitting}. {\it Right panel:}  Same as left panel, but for the higher-redshift regime ($z \sim 9-16$) derived from the intrinsic UV luminosity - halo mass relation, and assuming an escape fraction of  $0.15 \pm 0.05$.}
    \label{fig:laelumfunc}
\end{figure}

We employ the abundance matching technique to infer the Lyman-$\alpha$ luminosity as a function of host halo mass and its evolution with redshift. To do this, we use the observed  luminosity functions of Lyman-$\alpha$ emitters compiled over $z \sim 3.1-7.3$ from the Subaru/{\it XMM-Newton}, HSC and Cosmic Evolution Survey (COSMOS) fields \citep{ouchi2008, Ouchi10, konno2014, itoh2018}. We estimate the halo masses corresponding to the Lyman-$\alpha$ luminosities $L_{\rm Lya}$ and redshifts $z$, by solving the equation
\begin{equation}
  \int_{M_h (L_{\rm Lya})}^{\infty} \frac{dn}{ d \log_{10} M'} \ d \log_{10} M' = \int_{L_{\rm Lya}}^{\infty} \phi(L_{\rm Lya}) \ d \log_{10} L_{\rm Lya}
  \label{abmatchlae}
\end{equation}
for the observed $L_{\rm Lya}(M_h)$ (assumed monotonic) as a function of dark matter halo mass $M_h$, and using the Sheth-Tormen \citep{sheth2002} form of the comoving number density of dark matter haloes per unit logarithmic halo mass, ${dn}/{ d \log_{10} M_h}$. The procedure advocates a double power law form for the fit over $z \sim 3-7$, with the best-fitting parameters summarized in Table \ref{table:constraints}. The resulting Lyman-$\alpha$ luminosity to host halo mass relation is shown by the solid lines in the left panel of Fig. \ref{fig:laelumfunc}, with the shaded areas denoting the uncertainties inferred
from the Lyman-$\alpha$ luminosity functions. The fitting form is shown by the dashed lines in the same figure, { with the fit uncertainties shown by their enclosing shaded regions.  The fitting form is found to provide a good approximation to the observations within the errors, except for the lowest redshift slice where it deviates by about 0.3 dex for $L_{\rm Lya} < 10^{42}$ ergs/s}.

\begin{table}
\begin{tabular}{lll}
{\bf \large Summary of model parameters}  & & \\
\hline
&&\\

     {\large {\bf Observed LAE Luminosity - Halo mass} } &   &  \\
     \\
      {\underline{Fitting functions (all masses in $M_{\odot}$, luminosities in { $10^{31}$} ergs/s):}} &  & \\
     && \\
$L_{\rm LAE}(M,z) = 2N_1 M [(M/M_1)^{-\beta} + (M/M_1)^{\gamma}]^{-1} $ & & \\ 
  $z \sim 3.1$  & & \\
     
  {\underline{Parameter values:}} &  & \\
$M_1 = (1.22 \pm 0.18) \times 10^{12}$; $N_1 = (3.89 \pm 0.03) \times 10^{-1}$;
&& \\
 $\beta = -0.506 \pm 0.004$ ; $\gamma = -1.03 \pm 0.29$ &&\\
  &&\\
  
  $z \sim 3.7$  & & \\
     
  {\underline{Parameter values:}} &  & \\
$M_1 = (1.23 \pm 0.09) \times 10^{12}$; $N_1 = (5.88 \pm 0.02) \times 10^{-1}$;
&& \\
 $\beta = -0.523 \pm 0.003$ ; $\gamma = -0.75 \pm 0.11$ &&\\
  &&\\
  
  $z \sim 5.7$   & & \\
    
  {\underline{Parameter values:}} &  & \\
$M_1 = (7.37 \pm 0.68) \times 10^{11}$; $N_1 = (2.27\pm 0.04)$;
&& \\
 $\beta = 0.36 \pm 0.05$ ; $\gamma = 0.50 \pm 0.007$ &&\\
  &&\\
  
  $z \sim 6.6$   & & \\
    
  {\underline{Parameter values:}} &  & \\
$M_1 = (1.12 \pm 0.11) \times 10^{12}$; $N_1 = (3.02 \pm 0.06)$;
&& \\
 $\beta = 0.26 \pm 0.03$ ; $\gamma = 0.56 \pm 0.009$ &&\\
  &&\\
  $z \sim 7.0$  & & \\
      
  {\underline{Parameter values:}} &  & \\
$M_1 = (6.99 \pm 0.67) \times 10^{11}$; $N_1 = (3.62\pm 0.06)$;
&& \\
 $\beta = 0.35 \pm 0.05$ ; $\gamma = 0.54 \pm 0.009$ &&\\
  &&\\
  $z \sim 7.3$  & & \\
  {\underline{Parameter values:}} &  & \\
$M_1 = (3.65 \pm 0.54) \times 10^{11}$; $N_1 = (2.15\pm 0.15)$;
&& \\
 $\beta = -0.54 \pm 0.007$ ; $\gamma = -0.34 \pm 0.12$ &&\\
\noindent\rule{\linewidth}{0.1pt}
 &&\\

\end{tabular}
\caption{Best-fitting functional form describing the Lyman-$\alpha$ luminosity to host dark matter halo mass, inferred from abundance matching the observed Lyman-$\alpha$ luminosity functions over $z \sim 3-7$ to the Sheth-Tormen form of the halo mass function.} 
\label{table:constraints}
\end{table}

\begin{table}
\begin{tabular}{lll}
{\bf \large Summary of model parameters}  & & \\
\hline
&&\\

     {\large {\bf Observed LAE Luminosity - Halo mass} } &   &  \\
     \\
      {\underline{Fitting functions (all masses in $M_{\odot}$, luminosities in ergs/s):}} &  & \\
     && \\
$L_{\rm LAE}(M,z) =  M_1 [(M/10^9 M_{\odot})^{\beta}] $ & & \\ 
  $z \sim 9.0$  & & \\
     
  {\underline{Parameter values:}} &  & \\
$M_1 = (2.25 \pm 0.45) \times 10^{37}$; $\beta = 2.00 \pm 0.03$;
&& \\
  
  $z \sim 10.5$  & & \\
     
  {\underline{Parameter values:}} &  & \\
$M_1 = (1.05 \pm 0.16) \times 10^{38}$; $\beta = 1.91 \pm 0.02$;
&& \\
 
  $z \sim 12.0$   & & \\
    
  {\underline{Parameter values:}} &  & \\
$M_1 = (5.51 \pm 0.53) \times 10^{37}$;$\beta = 2.44 \pm 0.02$;
&& \\

  $z \sim 13.25$   & & \\
    
  {\underline{Parameter values:}} &  & \\
$M_1 = (2.99\pm 0.33) \times 10^{38}$; $\beta = 2.01 \pm 0.02$;
&& \\
    
      $z \sim 16.0$   & & \\  
  {\underline{Parameter values:}} &  & \\
$M_1 = (1.07 \pm 0.12) \times 10^{39}$; $\beta = 2.79 \pm 0.03$;
&& \\
 
\noindent\rule{\linewidth}{0.1pt}
 &&\\

\end{tabular}
\caption{Best-fitting functional form for of the Lyman-$\alpha$ luminosity - dark matter halo mass relation over $z \sim 9-16$, derived from the intrinsic UV luminosity functions and assuming an escape fraction of 15\%. Note the significantly different power law index compared to the lower redshift relations in Table \ref{table:constraints}.} 
\label{table:constraintshighz}
\end{table}

\begin{figure}
    \centering
    \includegraphics[width = 0.47\columnwidth]{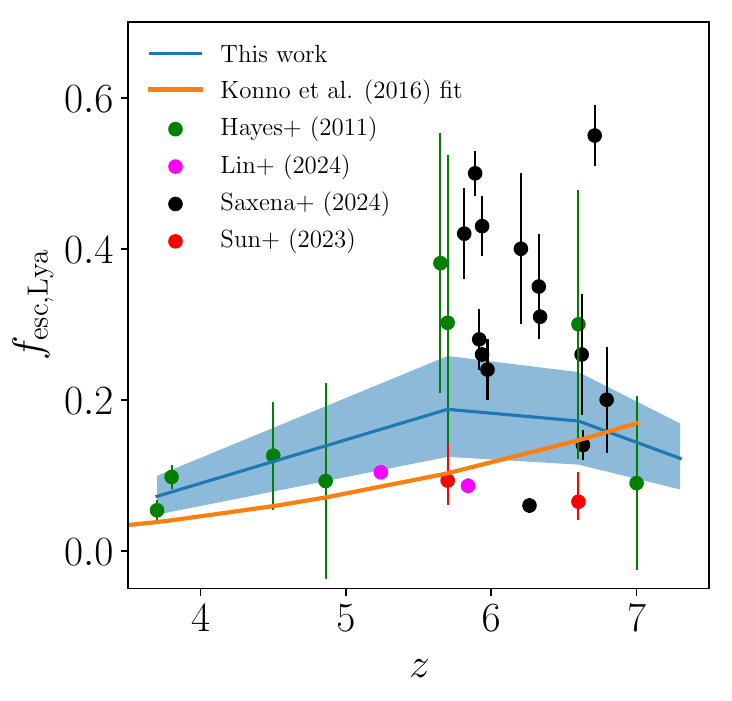}  \includegraphics[width = 0.47\columnwidth]{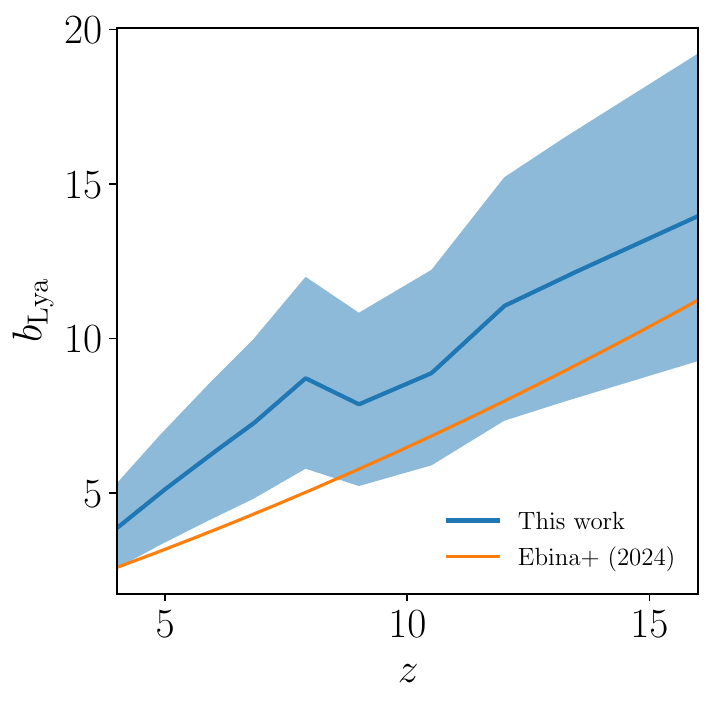} 
    \label{fig:fesc}
    \caption{{\it Left panel}: Evolution of escape fraction from the observed LAE luminosity and that derived from the dust-corrected UV luminosity. {\it Right panel:} Evolution of the bias of Lyman-$\alpha$ emission, compared to the fitting form of \citep{ebina2024} for a limiting flux of $10^{-17}$ erg/s/cm$^{-2}$.}
\end{figure}

We can also calculate the expected intrinsic Lyman-$\alpha$ luminosities of galaxies over the same redshift range as above. { This is done by using their observed star formation rate densities (SFRD),  measured at redshifts $z < 8$ from the HST UV luminosity functions compiled in \citep{bouwens2015}.}
The  intrinsic Lyman-$\alpha$  luminosity as a function of host halo mass can be derived from these densities by using the relation \citep{ouchi2020, konno2016}:
\begin{equation}
\rho(L_{\rm Lya} ({\rm int}))/({\rm ergs \ s^{-1} Mpc^{-3}}) = 1.1  \times 10^{42} \left(\frac{\rm SFRD}{\rm M_{\odot} yr^{-1} Mpc^{-3}}\right)
\label{uvtointlya}
\end{equation}
{ The above equation is derived from the combination of the H-$\alpha$ luminosity - SFR relation under  a constant ionizing photon production rate \citep{kennicutt1998} and the case B approximation \citep{brocklehurst1971}.}
We use these intrinsic  Lyman-$\alpha$ luminosity densities along with the observed ones inferred over the same redshift range by Ref. \citep{konno2018} from the Subaru surveys, to calculate the escape fraction (defined as the ratio of the observed to intrinsic luminosity density) as a function of redshift in the low-$z$ regime. As expected, its value and evolution are consistent with observational estimates. This is shown in the left panel of Fig. \ref{fig:fesc}, along with (i) the data points of \citep{hayes2011} inferred from older compilations of UV and Lyman-$\alpha$ luminosity densities \citep[e.g.,][]{iye2006, dawson2007, bouwens2009a} over the same redshift regime, (ii) the fitting form $f_{\rm esc} = 5 \times 10^{-4} (1+z)^{2.8}$ derived by \citep{konno2016},  (iii) the values of escape fraction inferred by the JWST FRESCO/NIRCam at the same redshifts \citep{lin2024}
{ and (iv) the escape fraction of Lyman-$\alpha$ emitters over the $z \sim 5.8-8$ redshift range  measured  from the JADES survey \citep{Saxena2023},  using their Balmer emission lines with an  inferred dust attenuation. 
While the various observations show a large scatter, the results indicate a gradual decrease of Lyman-alpha escape fractions with redshift, consistently with decreasing sizes of ionized bubbles in the IGM.
The observational trend suggests a Lyman-$\alpha$ escape fraction of $f_{\rm esc} \sim 0.1 - 0.2$  at the higher redshift regime.}

We can now find the expected Lyman-$\alpha$ luminosities for the higher redshift regime ($z > 10$) by { scaling the ultraviolet luminosity functions observed by JWST, according to the relation connecting UV luminosity to star formation rate \citep{madau1998}:
\begin{equation}
L_{\rm UV}/(\rm ergs  \ s^{-1} \ Hz^{-1}) = 8 \times 10^{27} \left(\frac{\rm SFR}{\rm M_{\odot} yr^{-1}}\right)
\end{equation}
used along with \eq{uvtointlya} above (expressed in the form $L_{\rm Lya} ({\rm int})$ = $1.1  \times 10^{42} \ {\rm SFR} $,  with $L_{\rm Lya}$(int) in units of ergs s$^{-1}$ and SFR in $M_{\odot}$ yr$^{-1}$, Ref. \citep{konno2016}). }

 Of course, these galaxies (at $z > 9$) are not expected to be observed directly in Lyman-$\alpha$, except in situations involving the presence of ionized bubbles \citep[e.g.,][]{Witstok-2024,  whitler2024}, but the results will be used to constrain the properties of Loeb-Rybicki haloes in the next section. We use the UV luminosity functions from the JWST early release NIRCam imaging data \citep{donnan2023, harikane2023b} and the abundance matching procedure employed in \citep{hploebjwst2023} which gives the corresponding halo masses. The intrinsic Lyman-$\alpha$ luminosities are calculated according to \eq{uvtointlya}, and then corrected for an escape fraction of $0.15 \pm 0.05$, as suggested by the lower-redshift findings. The resultant, expected Lyman-$\alpha$ luminosities as a function of host dark matter halo mass over $z \sim 9-16$ are plotted as the solid lines in the right panel of Fig. \ref{fig:laelumfunc}, with the shaded areas indicating the uncertainty on the escape fraction. As in the lower-redshift regime, these results can be fitted with a power law behaviour, whose parameters as a function of redshift are given in Table \ref{table:constraintshighz}.  { Similarly to the lower redshift range, the fit recovers the expected luminosity - host halo mass behaviour (within the uncertainties)  well   in the high luminosity-high halo mass range, $L_{\rm Lya} \gtrsim 10^{42}$ ergs/s, which is the regime of interest in the following sections.} In contrast  to Table \ref{table:constraints}, the results advocate a single power law, with a much steeper power law index and more aggressive evolution with redshift. 

{ The above analysis serves as an important extension of the corresponding results for the evolution of rest-frame UV luminosity functions, over the same redshift intervals (as summarized in, e.g.,  Refs. \citep{mashian2015,  hploebjwst2023}) to the Lyman-$\alpha$ regime.  It offers a useful comparison that highlights the rapid evolution of the intrinsic Lyman-$\alpha$ luminosity for $z > 9$, in sharp contrast to its behaviour over the $z \sim 3-8$ regime. While -- as emphasized earlier -- the Lyman-$\alpha$ luminosity function \textit{per se} is not directly observable in this case due to the Gunn-Peterson effect at $z \gtrsim 7$,  these values are crucial for estimating the expected luminosity of Loeb-Rybicki haloes around these early galaxies, as we will see in the following section.}

We also compute the clustering of LAEs over redshifts, which is given in terms of the effective bias, measured as \citep[e.g.,][]{khostovan2019}:
\begin{equation}
b^2 = \frac{\xi_{\rm gg}}{\xi_{\rm hh}}
\end{equation}
in which the numerator and denominator are the galaxy-galaxy and halo-halo correlation functions respectively.
In a halo model framework connecting dark matter halo mass to Lyman-$\alpha$ luminosity, the large scale value of the effective bias as a function of redshift can be modelled as \citep[e.g.,][]{hparaa2017, khostovan2019}:
\begin{equation}
b_{\rm Lya} (z) = \frac{\int_{M_{h,\rm min}}^{\infty} dM' (dn/dM') b_h (M',z) L_{\rm Lya} (M',z)}{\int_{M_{h,\rm min}}^{\infty} dM' (dn/dM') L_{\rm Lya} (M',z)}
\label{biasHInew}
\end{equation}
where the $dn/dM'$ denotes the dark matter halo mass function in the Sheth-Tormen form \citep{sheth2002} and the dark matter halo bias $b_h(M,z)$ follows, e.g., Ref. \cite{scoccimarro2001}. 
The large scale value of the effective bias has been estimated from 
{ an ensemble of observations of LAE clustering \citep{Khostovan19,Bielby16,Ouchi10,Shioya09,Hao18,Gawiser07,Guaita10} comprising the Slicing COSMOS 4K (SC4K) survey and archival NB497 imaging survey over $z \sim 2.5-5.8$, the VLT LBG survey at $z \sim 3$,  the Subaru/XMM-Newton Deep Surveys at $z \sim 2$,  $ z \sim 3.1 - 5.7$ and $z \sim 6.6$,  the Extended Chandra Deep Field-South (ECDF-S) at $z \sim 2.1$ and $z \sim 3.1$, and the COSMOS two square degree field at $z \sim 4.86$}. The results have been fit \citep{ebina2024} using a polynomial form:
\begin{equation}
    b(f_{\rm lim},z) = A(f_{\rm lim}) (1+z) + B(f_{\rm lim}) (1+z)^2 
    \label{bias_fit}
\end{equation}
where
\begin{align}
    A(f_{\rm lim}) &= 0.457-1.755(\log_{10}f_{\rm lim}+17)+0.720(\log_{10}f_{\rm lim}+17)^2 \\
    B(f_{\rm lim}) &= 0.012+0.318(\log_{10}f_{\rm lim}+17)+0.043(\log_{10}f_{\rm lim}+17)^2
\end{align}
in terms of the line flux limit of the survey, denoted by $f_{\rm lim}$ (measured in erg/s/cm$^{-2}$), and redshift $z$. The value of  $f_{\rm lim}$ directly translates into the minimum halo mass $M_{\rm min}(z)$ probed at each redshift, by inverting the relation in Table \ref{table:constraints} to solve for $M_{\rm min} = M_h(L_{\rm Lya, min}) \equiv M_h(4 \pi D_L(z)^2 f_{\rm lim})$, where $D_L(z)$ is the luminosity distance to redshift $z$.

The large scale bias calculated from the halo model framework, with a flux limit of $f_{\rm lim} = 10^{-17}$ erg/s/cm$^{-2}$ is plotted in the right panel of Fig. \ref{fig:fesc} by the blue solid line, with the shaded area indicating the associated  uncertainty. The best-fit from \eq{bias_fit} for the same line flux limit is shown by the solid orange line.  { While the fitting form of \cite{ebina2024} does not include error values, we can make an estimate of the expected scatter in the observations using the data points assembled from the surveys,  $\Delta b_{\rm Lya, obs} \approx 0.5$ (see also Fig. 1 of Ref. \cite{ebina2024}). The systematic offset between the orange curve and the blue shaded area may arise in part due to observational selection function effects in the surveys used in the fitting, over and above the flux limit of the measurements.} The curves are consistent within uncertainties,  validating the utility of the abundance matching procedure. \footnote{Abundance matching to the UV luminosity function is also expected to be valid over this redshift interval \citep{lee2009, mashian2015, hploebjwst2023} for galaxy ages of $> 100$ Myr for which several bursts of star formation are not expected.}

\begin{figure}
    \centering
    \includegraphics[width = 0.47\columnwidth]{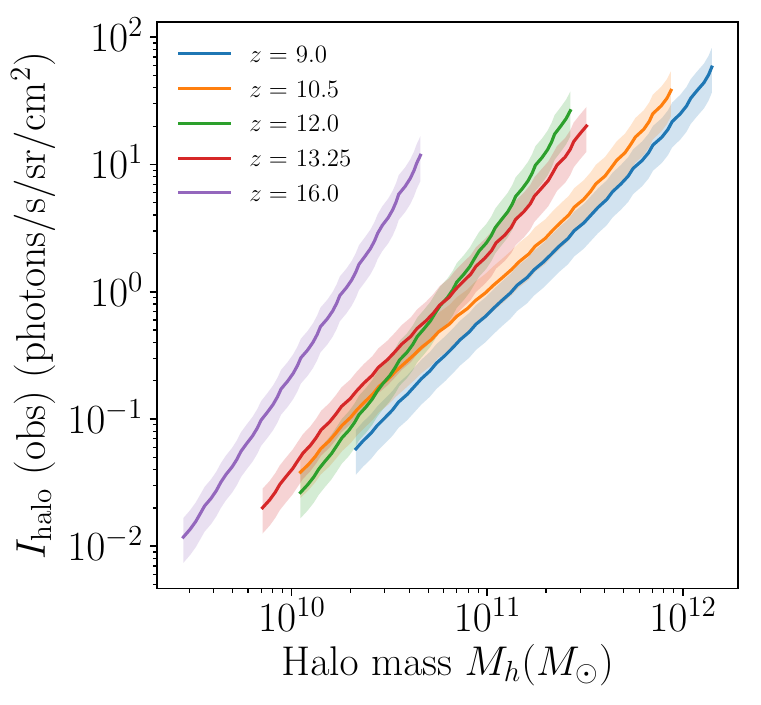}   \includegraphics[width = 0.47\columnwidth]{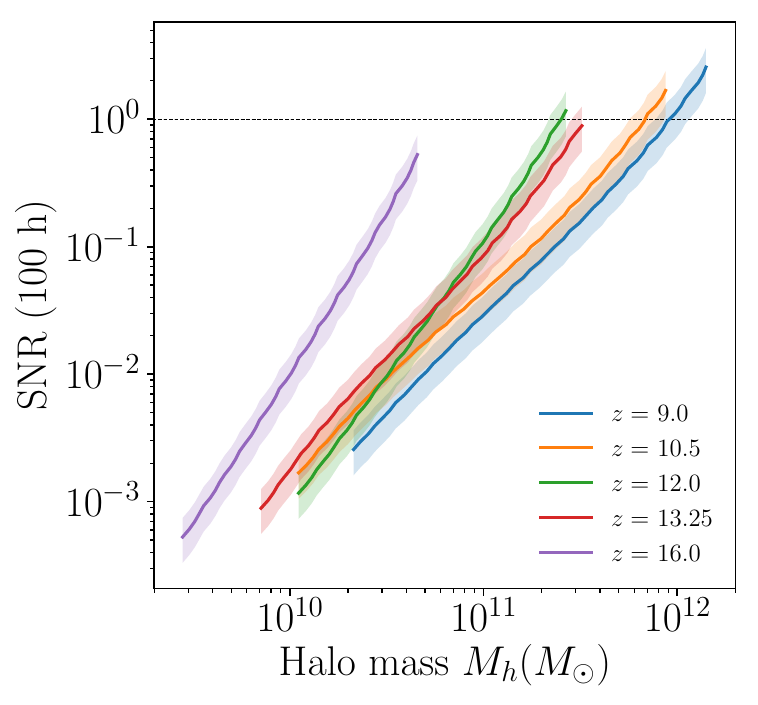} 
  
    \caption{{\it Left panel}: Observed central intensity of Loeb-Rybicki haloes at various redshifts, as a function of halo mass. {\it Right panel}: Signal-to-noise ratio (SNR) for detecting the Loeb-Rybicki haloes, plotted as a function of dark matter halo mass for 100 hours of JWST observations. { An SNR of unity is indicated by the black dashed line.}}
      \label{fig:ihalolumobs}
\end{figure}

\section{Detectability of Loeb-Rybicki haloes at $z >$ 10}
\label{sec:loebrybicki}

Having a developed data-driven forms to describe the expected Lyman-$\alpha$ emission from dark matter haloes over the $z \sim 9-16$ redshift range, we investigate the prospects for detecting Loeb-Rybicki haloes around the LAEs at these redshifts. The equation of radiative transfer in a uniform, expanding medium can be used to predict the observed surface brightness of the Lyman-$\alpha$ radiation that scatters around an LAE, assumed to be an isotropic and steady source of photons \citep{loeb1999}.\footnote{This ignores the ionized bubble around the LAE galaxy. This is a good approximation considering that the Loeb-Rybicki halo is expected to be larger than the ionized region, consistently with the results of most simulations and current JWST observations at lower redshifts \citep{lu2024, Witstok-2024, witstok2024a}.} It can be shown that the profile of the intensity distribution in such a halo of scattered radiation around a source of steady emission rate $\dot{N}_{\alpha}$ of Lyman-$\alpha$ photons, located at redshift $z_s$ is given by:
\begin{equation}
I(p)  = \frac{\tilde{I}(\tilde{p})}{(1 + z_s)^3}  \frac{\dot{N_{\alpha}}}{r_*^2}
\label{ihalop}
\end{equation}
In the above equation, $\tilde{I}(\tilde{p})$ is a dimensionless profile function{\footnote { See Fig. 1 of Ref.  \cite{loeb1999} for the shape of the profile as a function of the (dimensionless) impact parameter, as derived from a Monte Carlo calculation.}},  expressed as a function of an equivalent dimensionless impact parameter $\tilde{p}$. 

The maximum of the intensity occurs at zero impact parameter, and its value is given by substituting
$\tilde{I}_{\rm halo}(0) = 0.2$  into \eq{ihalop}. Here, $r_*$ is a characteristic scale, given as $\beta/\alpha^2$ where $\alpha$ and $\beta$ are defined as:
\begin{eqnarray}
\alpha &=& H (z_s) \nu_0/c =  2.7 \times 10^{-13} \ h \ (\Omega_m (1+ z_s)^3 + \Omega_{\Lambda})^{1/2} \ {\rm cm^{-1} Hz}; \ \nonumber \\
\beta &=& \left(\frac{3 c^2 \Lambda_{\alpha}^2}{32 \pi^3 \nu_0^2}\right) n_{\rm HI} = 1.5 \ \Omega_b \ h^2 (1+z_s)^3 \  {\rm cm^{-1} Hz^2}
\end{eqnarray}
The above two quantities are described in terms of fundamental constants, the speed of light $c$, and with $\Lambda_{\alpha} = 6.25 \times 10^8$ s$^{-1}$ being the spontaneous decay rate from the 2$p$ to 1$s$ state of neutral hydrogen. Here, $\nu_0 = 2.47 \times 10^{15}$ Hz is the Lyman-$\alpha$ rest frequency, and $H(z_s)$ is the Hubble parameter at the source redshift, $z_s$. The term $n_{\rm HI}$ denotes the number density of neutral hydrogen in the surrounding IGM, which is expressible in terms of the current baryon number density parameter, $\Omega_b$ in a uniform, dust free surrounding medium as $n_{\rm HI} = \Omega_b \rho_{c,0} (1 - Y_{\rm He}) (1 + z_s)^3/m_p $, where $Y_{\rm He} = 0.24$ is the helium fraction by mass, and $m_p$ is the mass of the hydrogen atom.
This leads to the value of $r_*$ given by
\begin{equation}
r_* = 2.1 \times 10^{25} \ {\rm cm} \frac{(\Omega_b/\Omega_m)}{1 + (\Omega_{\Lambda}/\Omega_m)(1+z_s)^{-3}}
\end{equation}

Given the luminosity as a function of halo mass for different redshifts (Fig. \ref{fig:laelumfunc}), we can derive the emission rate of Lyman-$\alpha$ photons per second as $\dot{N}_{\alpha} = L_{\rm Lya}/(h \nu_0)$. This quantity is plotted for the $z \sim 9-16$ redshift regime in the right-hand side axis of the right panel of Fig. \ref{fig:laelumfunc}. The peak emission rate reaches values of a few $10^{54}$ photons/s at the highest redshifts, about an order of magnitude greater than that observed at lower redshifts \citep{dey1998, loeb1999}. The resultant values of the observed central intensity, $I_{\rm halo} \equiv I(0)$, in photons/s/sr/cm$^{-2}$ in this redshift regime are plotted in the left panel of Fig. \ref{fig:ihalolumobs} as a function of host dark matter halo mass. 

We also forecast the detectability of these haloes by JWST. The detection is possible as long as the brightness of the halo exceeds the fluctuation noise of the sky background. Given that the surface brightness of the halo is expected to be constant over a scale 0.1$r_*$, the typical angular size of the halo is given approximately by $s = 0.1 r_*/D_A(z_s)$, where $D_A (z_s)$ is the angular diameter distance to the halo. This takes values $s \sim 24'' - 37''$ over the redshift range $z_s \sim 9-16$. 

We assume the sky brightness of JWST, which acts as an effective noise, to be given by $I_{\rm 0,sky} = 0.20 \ {\rm MJy/sr}$. This value is the median for the low-background zodiacal light emission \citep{rigby2023} and is a conservative estimate at $\lambda \sim 2 \mu m$, where the background can reach values of 0.15-0.18 MJy/sr. It is expected to be roughly constant over the 1.2 - 2 $\mu$m range, of interest for the present results.
The observed full width at half maximum of the Lyman-$\alpha$ line is found to be given by $\Delta \nu \sim \nu_*/(1+z_s)$, in terms of a frequency normalization scale $\nu_*$, defined by equating the optical depth from the source to the observer to unity:
\begin{equation}
\nu_* \equiv \beta/\alpha  = 5.6 \times 10^{12} \ h \ \Omega_b \ (1+z_s)^{3} \ [\Omega_m (1+z_s)^3 + \Omega_{\Lambda}]^{-1/2} \ {\rm Hz}
\end{equation}
This condition leads to a required filter band of $\Delta \nu = 8.4 \times 10^{11}$ Hz at 1.34 $\mu$m (for emission from $z_s \sim 10$). If only the innermost core of the emission is to be observed, a filter width of half this value is found to be sufficient \citep{loeb1999}. The sky noise over the filter band is then given by $I_{\rm sky} = 0.5 \Delta \nu I_{\rm 0,sky}/ (h \nu_{\rm obs})$ photons/cm$^2$/s/sr. For the redshift range under consideration ($z_s \sim 9-16$), the numerical values range from $I_{\rm sky} =(6.7 - 14.9) \times 10^5$ photons/cm$^2$/s/sr.

The signal-to-noise ratio, for an observation time $t_{\rm obs}$, collecting area $A$ of the telescope and angular area $ \Omega_{\rm area}$ of the Lyman-$\alpha$ halo on the sky    is then given by:
\begin{equation}
{\rm SNR} = \frac{t_{\rm obs}^{1/2} A^{1/2} \Omega_{\rm area}^{1/2}}{I_{\rm sky}^{1/2}}I_{\rm halo} = \frac{t_{\rm obs}^{1/2} A^{1/2} \Omega_{\rm area}^{1/2}}{I_{\rm sky}^{1/2}} \left(\frac{0.2 \dot{N}_{\alpha} }{(1+z_s)^3 r_*^2}\right)
\end{equation}
Given the collecting area of 25 m$^2$  for JWST, and $\Omega_{\rm area} \approx s^2$ with the angular size $s$ defined as $s = 0.1 r_*/D_A(z_s)$,  leads to the signal-to-noise ratio for fiducial 100 hour JWST observations, as a function of halo mass. This is plotted on the right panel of Fig. \ref{fig:ihalolumobs}. For the highest luminosities, individual haloes are detectable at the few sigma level for { $z \lesssim 9$}, while stacking about ten haloes can lead to detections at all redshifts.

We also calculate the spectral resolution (in km/s) required for observation of these haloes as a function of redshift. This is given by
\begin{equation}
\Delta v = \frac{0.5 c \Delta \nu }{\nu_{\rm obs}} = \frac{0.5 \ c \ \nu_* \ }{\nu_{0}} \approx 340 \ {\rm km/s} \  [\Omega_b  \ h \ (\Omega_m (1+ z_s)^{-3} + \Omega_{\Lambda} (1+ z_s)^{-6})^{-1/2}]
\end{equation}
For our fiducial cosmology, we have $\Delta v \sim 700-1500$ km/s over the redshift range under consideration.
The spectral resolution is $R \equiv c / \Delta v \sim 200-500$, which is also achievable with JWST \citep[e.g.,][]{napolitano2024, witstok2024a}.

\section{Conclusions}
\label{sec:conclusions}
We have compiled the recent measurements of the Lyman-$\alpha$ luminosity functions over $z \sim 3-7$ and used them to infer the evolution of the Lyman-$\alpha$ luminosity with dark matter halo mass. We have also used the dust-corrected UV luminosity functions to infer the intrinsic Lyman-$\alpha$ luminosities  of galaxies at these and higher redshifts. The lower redshift results indicate an escape fraction of 10-20\%, in line with recent and older observational results. The clustering of LAEs across redshifts is also consistent with that expected from fitting to observations. 

Applying the inferred escape fractions to the intrinsic Lyman-$\alpha$ luminosities derived from the UV luminosity functions at $z \sim 9-16$, gives us the expected evolution of the Lyman-$\alpha$ luminosity to host halo mass at these redshifts. While individual galaxy detection is difficult due to the large neutral hydrogen fraction expected around these early LAEs, we have  investigated how the Loeb-Rybicki Lyman-$\alpha$ haloes, produced by intergalactic scattering around the LAEs could be detectable by future JWST observations. The shift in frequency as a result of scattering allows these haloes to be detectable. It is found that for Lyman-$\alpha$ luminosities $> 10^{43}$ ergs/s,  individual haloes may be detectable up to { $z \lesssim 9$}, while stacking $\sim 10$ LAEs can lead to a detection of the signal out to $z \sim 16$. The angular size of the haloes is expected to lie in the $24''-37''$ range, with the detection requiring a  spectral resolution of a few hundred to a thousand km/s. 

Detection of Loeb-Rybicki haloes would map the neutral IGM independently of 21-cm surveys. Overdense filaments would show up as Lyman-$\alpha$ brightness enhancements.  Given that 65 LAEs  have been recently identified by JWST  \cite{napolitano2024} at lower redshifts, these findings promises good prospects for probing the earliest, as-yet elusive sources of reionization via their surrounding haloes.  { The integrated version of performing stacking is to apply the technique of intensity mapping \citep[IM; for a review see, e.g., Ref.][]{kovetz2019}, i.e. making a  statistical measurement of the clustering signal coming from all galaxies above a particular flux limit.  The resultant ‘stacked’ intensity of Loeb-Rybicki haloes at a particular redshift can be modelled by using a framework that connects their Lyman-alpha emissivities to the masses of their host dark matter haloes. The integrated signal is found to be measurable to a significance of a few tens of standard deviations out to $z \sim 16$ (for more details, see Ref. \citep{hploebimlya2024}) both in auto-correlation (correlating the signal with itself, measurable with the JWST, the upcoming SPHEREx \citep{dore2014}, and the planned Cosmic Dawn Intensity Mapper \citep[CDIM; Ref.][]{cooray2019}), as well as in cross-correlation  with the 21 cm emission from HI in the IGM, detectable by the Square Kilometre Array (SKA) and its pathfinder, the Murchison Widefield Array (MWA). 
} Apart from stacking, the gravitational clustering of galaxies around the primary LAE \citep[e.g.,][]{witstok2024a} may contribute to enhancing the signal. The scattered Lyman-$\alpha$ radiation is further expected to be highly polarized \citep{rybicki1999, masribas20} which may lead to enhanced prospects for its observation.
Refinements to the assumed isotropic model of Lyman-$\alpha$ emission, modifications to the geometry of emission and the effects of possible infall \citep[e.g.,][]{verhamme2008, smith2017, behrens2014, gronke2024} as captured by simulations, will be useful to increase the precision of forecasts and the interpretation of upcoming results from the JWST.  { These -- as well as a more detailed stacking analysis with observational data by summing the kernels of the Lyman-$\alpha$ haloes around individual galaxies in images --  will be the subject of future work.}

\section*{Acknowledgements}

{ We thank Richard Ellis for useful comments on the manuscript,  Sam Kramer for pointing out a typo in the caption of Table 1, and the referee for a detailed and helpful report.}
HP's research was supported by the Swiss National Science Foundation via Ambizione Grant PZ00P2\_179934. The work of AL was partially supported by the Black Hole Initiative at Harvard 
University, which is funded by grants from the JTF and GBMF.

\def\aj{AJ}                   
\def\araa{ARA\&A}             
\def\apj{ApJ}                 
\def\apjl{ApJ}                
\def\apjs{ApJS}               
\def\ao{Appl.Optics}          
\def\apss{Ap\&SS}             
\def\aap{A\&A}                
\def\aapr{A\&A~Rev.}          
\def\aaps{A\&AS}              
\def\azh{AZh}                 
\def\baas{BAAS}
\def\jcap{JCAP}
\def\jrasc{JRASC}             
\def\memras{MmRAS}
\def\na{New Astronomy}
\def\nat{Nature}
\def\mnras{MNRAS}             
\def\pra{Phys.Rev.A}          
\def\prb{Phys.Rev.B}          
\def\prc{Phys.Rev.C}          
\def\prd{Phys.Rev.D}          
\def\prl{Phys.Rev.Lett}       
\def\pasp{PASP}    
\def\pasa{PASA}               
\def\pasj{PASJ}
\def\physrep{Phys. Repts.}
\def\qjras{QJRAS}             
\def\skytel{S\&T}             
\def\solphys{Solar~Phys.}     
\def\sovast{Soviet~Ast.}      
\def\ssr{Space~Sci.Rev.}      
\def\zap{ZAp}                 
\let\astap=\aap
\let\apjlett=\apjl
\let\apjsupp=\apjs

\bibliography{mybib, main, biblio, refs}{}
\bibliographystyle{JHEP}

\label{lastpage}
\end{document}